\documentclass{ws-ijmpd}

\begin{document}

\title{Fluid sphere: stability problem and dimensional constraint}

\author{FAROOK RAHAMAN}
\address{Department of Mathematics, Jadavpur
University, Kolkata 700 032, West Bengal, India\\
rahaman@iucaa.ernet.in}

\author{ANIRUDH PRADHAN}
\address{Department of Mathematics, GLA University, Mathura 281406, U.P.,
India\\pradhan@iucaa.ernet.in}

\author{NASR AHMED}
\address{Mathematics Department, Deanery of Academic Services,
Taibah University, Saudi Arabia and Astronomy Department, National
Research Institute of Astronomy \& Geophysics, Helwan, Cairo,
Egypt
\\abualansar@gmail.com}

\author{SAIBAL RAY}
\address{Department of Physics, Government College of
Engineering \& Ceramic Technology, Kolkata 700 010, West Bengal,
India\\saibal@iucaa.ernet.in}

\author{BIJAN SAHA}
\address{Laboratory of Information Technologies, Joint Institute for
Nuclear Research, 141980 Dubna, Moscow region,
Russia\\bijan@jinr.ru}

\author{MOSIUR RAHAMAN}
\address{Department of Mathematics, Meghnad Saha Institute of Technology,
Kolkata 700150, West Bengal, India\\mosiurju@gmail.com}

\maketitle

\begin{history}
\received{Day Month Year} \revised{Day Month Year} \comby{Managing
Editor}
\end{history}

\begin{abstract}
We study different dimensional fluids inspired by noncommutative
geometry which admit conformal Killing vectors. The solutions of
the Einstein field equations examined specifically for five
different set of spacetime. We calculate the active gravitational
mass and impose stability conditions of the fluid sphere. The
analysis thus carried out immediately indicates that at
$4$-dimension only one can get a stable configuration for any
spherically symmetric stellar system and any other dimensions,
lower or higher, becomes untenable as far as the stability of a
system is concerned.
\end{abstract}


\section{Introduction}

Higher dimensional spacetime configuration is particularly
important in the studies of the early phase of the Universe. It
plays a significant role to describe not only the universe in
early stage of its evolution but also some unobservable phenomena
in the physical universe.

It is interesting to point out that Barrow \cite{Barrow1983} was
the first to investigate the role of spacetime dimensions in
determining the form of various physical laws and constants of
Nature. Basically he \cite{Barrow1983} employed the concept of
fractal dimension under Kaluza-Klein theories obtained by
dimensional reduction from higher dimensional gravity or
supergravity theories. It has therefore been argued that the
presently observed $4D$ spacetime is the compactified form of
manifold with higher dimensions and as such in the arena of grand
unification theory and also in superstring theory this kind of
self-compactification idea of multidimensions have been invoked by
different researchers \cite{Schwarz1985,Weinberg1986}. There are
also several other research articles available in the literature
in connection to higher spatial dimensions some of which can be
consulted in the following
Refs.~\cite{Liu2000,Dvali2000,Dvali2003,Iorio2005a,Iorio2005b,Iorio2005c,Iorio2006a,Iorio2006b,Kagramanova2006,Rahaman2009,Andrews2013,Renzetti2013}.

In cosmology, the Kaluza-Klein inflationary theory with higher
dimensions, it is believed that extra dimensions are reducible
specially to four dimension which was associated with some
physical processes. Ishihara \cite{Ishihara1984} and later on
Gegenberg and Das \cite{Gegenberg1985} have shown that within the
Kaluza-Klein inflationary scenario of higher dimension a
contraction of the internal space causes the inflation of the
usual space. It is argued by Iban{\`e}z and Verdaguer
\cite{Ibanez1986} that there are cases in FRW cosmologies where
the extra dimensions contract as a result of cosmological
evolution.

It has been observed that symmetries of geometrical as well as
physical relevant quantities of general relativity, known as
collineations and conformal Killing vectors (CKV), are most useful
to facilitate generation of exact solutions to the Einstein field
equations \cite{Rahaman2013b}. Since it is a key ingredient of the
methodology of the present paper so we would like to add here a
few more sentences in connection to collineations and conformal
Killing vectors. We note that a great details on conformal
collineations (a generalization of conformal Killing vector fields
and affine collineations and projective collineations) that
inherit the Ricci curvature (that is, preserve it up to a scalar
factor) have been analyzed on semi-Riemannian manifolds with
divergence-free Riemannian curvature tensors by Beem and
Duggal~\cite{Beem1994}. Apostolopoulos and
Tsamparlis~\cite{Apostolopoulos2001} argue that the proper Ricci
and Matter (inheritance) collineations are the (conformal) Killing
vectors of the generic metric which are not (conformal) Killing
vectors of the space-time metric and using this observation they
compute the Ricci and Matter inheritance collineations of the
Robertson-Walker space-times and we determine the Ricci and Matter
collineations without any further calculations. However, this
conformal Ricci and Matter Collineations technique has been
employed for an anisotropic fluid~\cite{Sharif2007}. There are
some other notable works on collineations and conformal Killing
vectors available in the following
Refs.~\cite{Duggal1986,Gleiser1992,Stephani2003,Tsamparlis2007,Ramzan2013}.

Therefore, in the present investigation following Yavuz et al.
\cite{Yavuz2005}, we have imposed the condition that the spacetime
manifold admits a CKV and thus tried to tackle the anisotropic
field equations in a suitably better way.

In connection to the anisotropic field equations of general
relativity the study of compact objects has been of ample interest
for a long time. It was argued long ago by Bowers and Liang
\cite{Bowers1917} that the effects of local anisotropy may have
important role for relativistic fluid spheres to attain
hydrostatic equilibrium in connection to maximum equilibrium mass
and surface redshift. Ruderman \cite{Ruderman1972} showed that in
the stellar interior the nuclear matter may have anisotropic
features at least in certain very high density ranges
($>10^{15}~gm/cm^3$) and thus advocated to treat the nuclear
interaction relativistically. Very recently, based on some
observed compact stars Kalam et al. \cite{Kalam2012} made an
extensive analysis to show the anisotropic behavior of the
samples.

Therefore, under the above theoretical background our motivation
is indeed to study stability problem and dimensional constraint in
connection to solutions of higher dimensional spherically
symmetric systems within the framework of noncommutative geometry.
The present investigation thus based on the following scheme:
After providing the basic field equations of Einstein in the Sec.
II, we seek the solutions under CKV in Sec. III for various
dimensions ranging from $3D$ to $11D$ spacetime. Sec. IV deals
with active gravitational mass where we get indication for minimum
potential energy at $4D$ spacetime. Therefore, in Sec. V we
continue our investigation by imposing equilibrium conditions and
arrive at the conclusion that higher dimensional spacetime is not
tenable for stability of the system. Sec. VI offers some
concluding remarks in favor of the results obtained.

\section{The Basic Field Equations of Einstein}
The spacetime metric describing a spherically symmetric system in
higher dimension is taken as
\begin{equation}
     ds^2=  - e^{\nu(r)} dt^2+ e^{\lambda(r)} dr^2+r^2 d\Omega_n^2,
   \label{Eq3}
   \end{equation}
where the line element ${d{\Omega}_n}^2$ on the unit $n$-sphere
is given by
\begin{eqnarray}
{d{\Omega}_n}^2 = d {{{\theta}_1} ^2} + {sin}^2 {\theta}_1 d
{{{\theta}_2} ^2}+ {sin}^2 {\theta}_1 {sin}^2 {\theta}_2 d
{{{\theta}_3} ^2} \nonumber\\ +..................
+\prod_{i=1}^{n-1} {sin}^2 {\theta}_i d{{\theta}_n} ^2.
\end{eqnarray}

The general energy momentum tensor which is compatible with static
spherically symmetry  as
\begin{equation}
T_\nu^\mu=  ( \rho + p_r)u^{\mu}u_{\nu} - p_r g^{\mu}_{\nu}+
            (p_t -p_r )\eta^{\mu}\eta_{\nu} \label{eq:emten}
\end{equation}
with $u^{\mu}u_{\mu} = - \eta^{\mu}\eta_{\mu} = 1$.

The Einstein equations (for the geometrized units $G=c=1$) are
\begin{equation}\label{E:Einstein1}
e^{-\lambda}
\left(\frac{n \lambda^\prime}{2 r} - \frac{n(n-1)}{2r^2}
\right)+\frac{n(n-1)}{2r^2}= 8\pi \rho,
\end{equation}

\begin{equation}\label{E:Einstein2}
e^{-\lambda}
\left(\frac{n(n-1)}{2r^2}+\frac{n
\nu^\prime}{2r}\right)-\frac{n(n-1)}{2r^2}= 8\pi p_r ,
\end{equation}

\begin{eqnarray}\label{E:Einstein3}
\frac{e^{-\lambda}}{2} \left[\frac{1}{2}(\nu^\prime)^2+
\nu^{\prime\prime} -\frac{1}{2}\lambda^\prime\nu^\prime +
\frac{(n-1)}{r}({\nu^\prime- \lambda^\prime})\right. \nonumber
\\ \left.+\frac{(n-1)(n-2)}{r^2}\right] -\frac{(n-1)(n-2)}{2r^2}=8\pi
p_t,
\end{eqnarray}
where $\rho$, $p_r$ and $p_t$ are respectively the energy density,
radial pressure and tangential pressure of the static fluid
sphere. Here $\prime$ over $\nu$ and $\lambda$ denotes partial
derivative w.r.t. radial coordinate $r$ only.

The energy density having a minimal spread Gaussian profile in
higher dimension is taken as \cite{Spallucci2008,Rahaman2013c}
\begin{equation}\label{density1}
\rho = \frac{m}{(4 \pi \theta)^{{(n+1)/2}}}~{exp}
\left(-\frac{r^2}{4\theta}\right),
\end{equation}
Here, $m$ is the total mass of the source which can be diffused
throughout a region of linear dimension $\sqrt{\theta}$ due to the
uncertainty and generally assumed to be closed to the Planck
length scale.

\section{The Solutions under Conformal Killing Vectors}
To find the exact solution, as is indicated in the introductory
part, we use the well known inheritance symmetry of the spacetime
as the symmetry under conformal Killing vectors
\cite{Pradhan2007,Rahaman2012a} given as:
\begin{equation}
                L_\xi g_{ik} =\xi_{i;k}+ \xi_{k;i} = \psi g_{ik},
         \label{CKV}
          \end{equation}
where $\psi$ is an arbitrary function of $r$.

 The above conformal Killing Eqs. (\ref{CKV}) provide the
 following set:
 \begin{eqnarray}
               \xi^1 \nu^\prime =\psi;~~\\
               \xi^{n+2}  = c_1 = ~constant;~~\\
               \xi^1  = \frac{\psi r}{2};~~\\
               \xi^1 \lambda ^\prime + 2 \xi^1 _{,1}   =\psi,
                \label{Eq3}
          \end{eqnarray}
where the subscript of comma denotes the partial derivative w.r.t.
$r$. Here in the second expression $\theta$'s are involved,
however being constant we have taken those as $c_1$.

These Eqs. (9) - (12) imply
\begin{equation}
               e^\nu  =c_2^2 r^2,
     \label{nu}
          \end{equation}

\begin{equation}
               e^\lambda  = \left(\frac {c_3} {\psi}\right)^2,
     \label{lam}
          \end{equation}

\begin{equation}
               \xi^i = c_1 \delta_{n+2}^i + \left(\frac{\psi
               r}{2}\right)\delta_1^i,
       \label{xi}
          \end{equation}
where $c_2$ and $c_3$ are integration constants. The above Eqs.
(\ref{nu}) - (\ref{xi}) contain all the characteristic features
derived from the existence of the conformal collineation.

Now, using the values of $\nu$ and $\lambda$ from Eqs. (\ref{nu})
and (\ref{lam}), we can solve the field equations for the given
energy density (\ref{density1}), i.e. we try to find out here the
unknowns $\psi$, $p_r$ and $p_t$ under different space-time with
varying value of $n$, the index of dimension. As a sample study we
only consider cases with $n=1$, $n=2$, $n=3$, $n=8$ and $n=9$
representing $3D$, $4D$, $5D$, $10D$ and $11D$ spacetime
respectively.

\subsection{$3D$ space-time ($n=1$)}
Starting this case of lower space-time with $n=1$, the metric
potential can be given by
\begin{equation}
\lambda(r)=\ln\left(\frac{\theta}{4m(2\theta
e^{-\frac{r^{2}}{4\theta}}-c_{1})}\right),
\end{equation}
where $c_1$ is an integration constant as mentioned earlier. Here
in the above expression $\theta$ in the numerator and as
multiplier in the denominator are im posed manually and can be
taken as unity for simplicity.

Here the pressures are taking the forms:
\begin{equation}
p_{r}=\frac{m(2\theta
e^{-\frac{r^{2}}{4\theta}}-c_{1})}{2\pi\theta r^{2}},
\end{equation}

\begin{equation}
p_{t}=-\frac{me^{-\frac{r^{2}}{4\theta}}}{4\pi\theta}.
\end{equation}

In a similar way, by applying the boundary conditions at $r=R$,
i.e. $p_R=0$, at once we get
\begin{equation}
c_{1}=2\theta e^{-\frac{R^{2}}{4\theta}}.
\end{equation}

\subsubsection{Matching conditions}

Our interior metric is
\begin{equation}
ds^{2}=-c_{2}^{2}r^{2}dt^{2}+e^{\lambda(r)}dr^{2}+r^{2}d\Omega_{n=1}.
\end{equation}
We match our interior solution to the exterior BTZ metric given by
\begin{equation}
ds^{2}=-(-M_{0}-\Lambda r^{2})dt^{2}+(-M_{0}-\Lambda
r^{2})^{-1}dr^{2}+r^{2}d\Omega_{n=1}.
\end{equation}

So the matching conditions yield the following results:
\begin{equation}
c_{1}=(M_{o}+\Lambda R^{2})\frac{\theta}{4m}+2\theta
e^{-\frac{R^{2}}{4\theta}},~~~
c_{2}=\frac{1}{R}\sqrt{-(M_{o}+\Lambda R^{2})}.
\end{equation}

If we now put the value of $c_1$ from Eq. (19), we get $M_0
+\Lambda R^2 =0$ which is clearly impossible. Therefore, three
dimensional fluid inspired by non commutative geometry admitting
conformal killing vector is not physically viable.

\subsection{$4D$ space-time ($n=2$)}
This is the usual $4D$ space-time and in this case the metric
potential can be given by
\begin{equation}
\lambda(r)=\ln\left[\frac{r\pi\sqrt{\pi\theta}}{-\pi\sqrt{\pi\theta}\left(r+c_{1}-2m~erf\left(\frac{r}{2\sqrt{\theta}}\right)\right)-
2mre^{-\frac{r^{2}}{4\theta}}\pi}\right],
\end{equation}
where $erf\left(\frac{r}{2\sqrt{\theta}}\right)$ is the error
function.

Therefore, the radial and tangential pressure parameters, $p_r$
and $p_t$, now are respectively taking the following forms:
\begin{equation}
p_{r}=\frac{-2r+6m~erf\left(\frac{r}{2\sqrt{\theta}}\right)-3c_{1}-\frac{6mr}{\sqrt{\pi\theta}}e^{-\frac{r^{2}}{4\theta}}}{8r^{3}\pi},
\end{equation}

\begin{equation}
p_{t}=\frac{\left(-\sqrt{\pi}\theta^{2}+mr^{2}\sqrt{\theta}e^{-\frac{r^{2}}{4\theta}}\right)}{8(r\theta)^{2}(\pi
\theta)^{3/2}}.
\end{equation}

At the boundary surface ($r=R$) pressure should be considered as
of vanishing order ($p_{r=R}=0$). Thus we get the value of the
constant $c_{1}$ as
\begin{eqnarray}
c_{1}=\frac{2}{3}\left[-R-3\frac{mR}{\sqrt{\pi\theta}}e^{-\frac{R^{2}}{4\theta}}+3m~erf\left(\frac{R}{2\sqrt{2}}\right)\right].
\end{eqnarray}

\subsubsection{Matching conditions} In the present case our metric:
\begin{equation}
ds^{2}=-c_{2}^{2}r^{2}dt^{2}+e^{\lambda(r)}dr^{2}+r^{2}d\Omega_{n=2}.
\end{equation}

On the other hand, $4D$ metric is Schwarzschild metric and can be
supplied as
\begin{equation}
ds^{2}=-\left(1-\frac{2M}{r}\right)dt^{2}+\left(1-\frac{2M}{r}\right)^{-1}dr^{2}+r^{2}d\Omega_{2}.
\end{equation}

So matching conditions provide us the following expressions for
the constant quantities:
\begin{eqnarray}
c_{1}= -2R+2M + 2m ~erf\left(\frac{R}{2\sqrt{\theta}}\right) -
\frac{2m R}{\sqrt{\pi \theta}}
e^{-\frac{R^{2}}{4\theta}},\nonumber\\~~~~~~~~~~~~~~~~~~~~~
c_2=\frac{1}{R}\sqrt{\frac{2M}{R}-1}.~~~~~~~~~~~~~~~~~~~~~
\end{eqnarray}

\subsection{$5D$ space-time ($n=3$)}
Let us now move towards higher dimension by choosing the value of
$n > 2$. The metric potential, the radial and tangential pressures
for this $5D$ space-time can respectively be given by
\begin{equation}
\lambda(r)=\ln\left[\frac{3r^{3}\pi^{2}\theta}{\pi^{2}\theta(3r^{2}+2c_{1})+2\pi
m(r^{2}+4\theta)e^{-\frac{r^{2}}{4\theta}}}\right],
\end{equation}

\begin{equation}
p_{r}=\frac{\pi
\theta(9r^{2}+4c_{1})+4m(r^{2}+4\theta)e^{-\frac{r^{2}}{4\theta}}}{8\pi^{2}r^{4}\theta},
\end{equation}

\begin{equation}
p_{t}=-\frac{-4\pi\theta^{2}+mr^{2}e^{-\frac{r^{2}}{4\theta}}}{16\pi^{2}r^{2}\theta^{2}}.
\end{equation}

At $r=R$, $p_r({r=R})=0$ and that gives
$c_{1}=-\frac{9}{4}R^{2}-\frac{m}{\pi
\theta}(R^{2}+4\theta)\exp(-\frac{R^{2}}{4\theta})$.

\subsubsection{Matching conditions} Our metric in this case is
\begin{equation}
ds^{2}=-c_{2}^{2}r^{2}dt^{2}+e^{\lambda(r)}dr^{2}+r^{2}d\Omega_{n=3}.
\end{equation}

Again, the $5D$ Schwarzschild metric is
\begin{equation}
ds^{2}=\left(1-\frac{8  M}{3\pi
r^{2}}\right)dt^{2}+\left(1-\frac{8 M}{3\pi
r^{2}}\right)^{-1}dr^{2}+r^{2}d\Omega_{3}.
\end{equation}

So matching conditions on the boundary immediately give us:
\begin{eqnarray}
c_{1}= \frac{3}{2}(R^{3}-R^2)-\frac{4 MR}{\pi}
-\frac{m}{4\theta}(R^2+4
\theta)e^{-R^{2}/4\theta},\nonumber\\~~~~~~~~~~~~~~~~~~~~~
c_{2}=\frac{1}{R}\sqrt{1-\frac{8 M}{3\pi
R^{2}}}.~~~~~~~~~~~~~~~~~~~~~
\end{eqnarray}

\subsection{$10D$ space-time ($n=8$)}
For the arbitrary choice of $n=8$ for higher dimensional case, we
get the following results:
\begin{eqnarray}
\lambda(r)=\ln(128r^{7}\pi^{5}\theta^{4}) \nonumber \\~~~~~~~
-\ln\left[32\theta^{4}\pi^{5}(4r^{7}+c_{1})-840\theta^{4}\pi^{2}m~erf\left(\frac{r}{2\sqrt{\theta}}\right)\right.
\nonumber \\~~~~~~
\left.+\sqrt{\pi\theta}me^{-\frac{r^{2}}{4\theta}}\left(\pi
r^{7}+14\theta^{3}\pi r^{5}+140\theta^{5}r^{3}+840\theta^{7}\pi
r\right) \right],~~~
\end{eqnarray}

\begin{eqnarray}
p_{r}=-\frac{1}{256r^{9}\pi^{6}}
\left(\pi^{5}(-2048r^{4}-288c_{1})\right.\nonumber \\
\left.+7560\pi^{2}
m~erf\left(\frac{r}{2\sqrt{\theta}}\right)\right)\nonumber \\
-\frac{\sqrt{\theta
\pi}me^{-\frac{r^{2}}{4\theta}}}{256r^{9}\pi^{6}\theta^{4}}\left(-9\pi
r^{7}\right.\nonumber \\
\left.-126\theta \pi r^{5}-126\theta^{2}\pi
r^{3}-7560\theta^{3}\pi r\right),
\end{eqnarray}

\begin{equation}
p_{t}=-\frac{-448\pi^{4}\theta^{\frac{1}{2}}+\sqrt{\pi}r^{2}me^{-\frac{r^{2}}{4\theta}}}{512r^{2}\pi^{5}\theta^{9/2}}.
\end{equation}

At $r=R$, $p_{r}=0$ and it provides
\begin{eqnarray}
c_{1}=-\frac{2048R^{7}}{288}+\frac{7560}{288\pi^{3}}m~erf\left(\frac{R}{2\sqrt{\theta}}\right)-
\nonumber
\\ \frac{\sqrt{\pi\theta}me^{-\frac{R^{2}}{4\theta}}}{288\pi^{4}\theta^{4}}
 \left(9R^{7}+126\theta R^{5}+1260\theta^{2}
R^{3}+7560\theta^{3} R\right).
\end{eqnarray}

\subsection{$11D$ space-time ($n=9$)}
We would like to study one more higher dimensional case with
$n=9$. In this chosen case of $11D$ space-time, the metric
potential and pressure parameters take the forms:
\begin{eqnarray}
\lambda(r)=-\ln\left[\frac{64c_{1}+288r^{8}}{288r^{8}} \right.
~~~~~~~~~~~~~~~~~~~~~~~~~~~~~~~~~~~~   \nonumber \\ \left.
+\frac{me^{-\frac{r^{2}}{4\theta}}(6144\theta^{4}+1536r^{2}\theta^{3}+192r^{4}\theta^{2}+16
r^{6}\theta+r^{8})}{r^{8}\theta^{4}\pi^{4}}\right],
\end{eqnarray}

\begin{eqnarray}
p_{r}=\frac{320c_{1}+2592r^{8}}{256\pi r^{10}}
~~~~~~~~~~~~~~~~~~~~~~~~~~~~~~~~~~~~  \nonumber \\
+\frac{me^{-\frac{r^{2}}{4\theta}}(30720\theta^{4}+7680r^{2}\theta^{3}+960r^{4}\theta^{2}+80
r^{6}\theta+5r^{8})}{256r^{10}\theta^{4}\pi^{5}},
\end{eqnarray}

\begin{equation}
p_{t}=-\frac{-1024\theta^{5}\pi^{4}+r^{2}me^{-\frac{r^{2}}{4\theta}}}{1024r^{2}\theta^{5}\pi^{5}}.~~~~~~~~~~~~~~~~~~~~~~~~~~~~~~~~~
\end{equation}

At $r=R$, $p_{r=R}=0$ and we get
\begin{eqnarray}
c_{1}=-\frac{2592R^{8}}{320}-\frac{me^{\frac{-R^{2}}{4\theta}}}{320\pi^{4}\theta^{4}}~~~~~~~~~~~~~~~~~~~~~~~~~~~~~~~~~
\nonumber\\ \times
(5R^{8}+80R^{6}\theta+960R^{4}\theta^{2}+7680R^{2}\theta^{3}+30720\theta^{4}).
\end{eqnarray}

\section{Active gravitational mass in various dimensions}
We apply the following relation to calculate active gravitational
mass in various dimensions \cite{Rahaman2006}
\begin{equation}
M(R)=\int_{0}^{R}\left[\frac{2\pi^{\frac{n+1}{2}}}{\Gamma(\frac{n+1}{2})}\right]
r^{n}\rho(r)dr.
\end{equation}

\small\begin{figure}
    \begin{center}
\includegraphics[scale=.4]{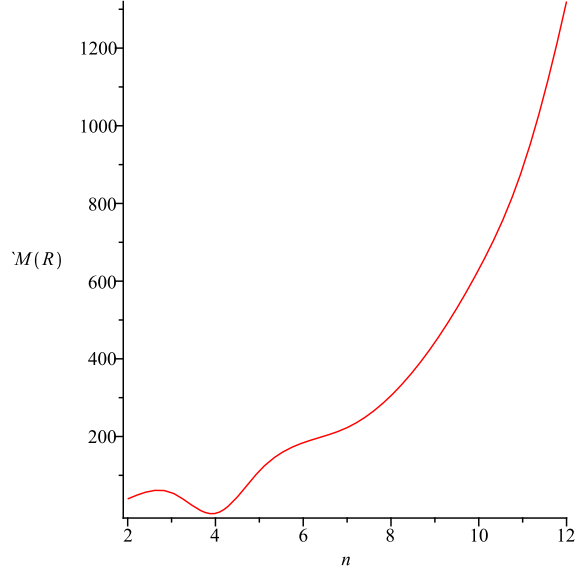}
    \end{center}
    \caption{A plot of $M(R)$ (as vertical axis) with spacetime dimensions $n$ (as horizontal axis) is drawn.
    In the figure specifically we have used the numerical values
$m=1.4~M_{\odot}$, $R=10$~km and $\theta=0.002$ to plot $M(R)$ vs
$n$ graph. It is evident from the Fig. 1 that the function $M(R)$
increases in a blowing up manner as the number of dimensions
increases except at the $4D$ spacetime which has a minimum value.}
    \label{rf}
\end{figure}\normalsize

Now before drawing any conclusion from the above expression for
mass, let us first see what is the lesson to be extracted from
this study for our understanding of possible properties of
different static and spherically symmetric fluid distribution.
This obviously can be done through a systematic survey of the
effect of higher dimensional spacetime for the different parameter
set that fit for the observed stars. We have collected data for
the masses of some compact objects, e.g. PSR J1614-2230 [$1.97 \pm
0.08~M_{\odot}$ \cite{Demorest2010}], PSR J1903+327 [$1.667 \pm
0.02~M_{\odot}$ \cite{Freire2011}], Vela X-1 [$1.77 \pm
0.08~M_{\odot}$ \cite{Rawls2011}], SMC X-1 [$1.29 \pm
0.05~M_{\odot}$ \cite{Rawls2011}] and Cen X-3 [$1.29 \pm
0.08~M_{\odot}$ \cite{Rawls2011}]. The corresponding radii have
been calculated by Takisa et al. \cite{Takisa2014} which
respectively are as follows: 10.30~km, 9.82~km, 9.99~km, 9.13~km
and 9.51~km (also see the Refs.
\cite{Kalam2012,Rahaman2012c,Hossein2012,Kalam2013} for other data
set for different compact stars).

Keeping in mind the above range of masses and radii, let us then
realistically consider the following specific physical
configuration of any compact star:\\ With the numerical values
$m=1.4~M_{\odot}$, $R=10$~km and $\theta=0.002$, from a straight
forward calculation, we get $M(R)=39.39$, $55.70$, $1.98$,
$111.40$, $222.81$, $630.22$, $891.27$, $2520.88$ for the
dimensions $2$, $3$, $4$, $5$, $7$, $10$, $11$ and $14$
respectively. It is very interesting to note from the Fig. 1 that
the function $M(R)$ increases in a blowing up manner as the number
of dimensions increases but it has only one minimum at the $4D$
spacetime. There is also a small hump visible at $3D$ spacetime.
This means that noncommutative geometry admitting conformal
Killing vectors with anisotropic fluid sphere permit only the
$4$-dimensional spacetime to make the spherically symmetric matter
distribution in stable equilibrium.

\section{Equilibrium: TOV equation for $4D$}
In view of the above result and discussion we are then in demand
of verifying the equilibrium features of the spherically symmetric
matter distribution. The active gravitational mass, for the
present case, can be given by
\begin{equation}
-M_{g}\frac{p_{r}+\rho}{r^{2}}e^{\frac{\lambda-\nu}{2}}-\frac{dp_{r}}{dr}+\frac{2}{r}(p_{t}-p_{r})=0,
\end{equation}
where
\begin{equation}
M_{g}=M_{g}(r)=\frac{1}{2}r^{2}e^{\frac{\nu-\lambda}{2}}
\frac{d\nu}{dr}.
\end{equation}

Now equilibrium of the spherical symmetric system requires the
following condition:
\begin{equation}
F_{g}+F_{h}+F_{a}=0,
\end{equation}
where $F_{g}$, $F_{h}$ and $F_{a}$ are respectively the
gravitational, hydrostatic and anisotropic forces.

Let us then provide the forces in action, i.e. gravitational,
hydrostatic and anisotropic, respectively as follows:
\begin{eqnarray}
F_{g}=-M_{g}\frac{p_{r}+\rho}{r^{2}}e^{\frac{\lambda-\nu}{2}}
~~~~~~~~~~~~~~~~~~~~~~~~~~~~~~~~~~\nonumber
\\=-\frac{1}{r}\left[\frac{-2r+6m~erf(\frac{r}{2\sqrt{\theta}})
-3c_1-\frac{6mr}{\sqrt{\pi \theta}}e^{-\frac{r^2}{4\theta}}}{8\pi
r^3} \right.  \nonumber \\
\left.+ \frac{m}{(4\pi
\theta)^{\frac{3}{2}}}e^{-\frac{r^2}{4\theta}}\right],
\end{eqnarray}

\begin{eqnarray}
F_{h}=\frac{dp_{r}}{dr}
~~~~~~~~~~~~~~~~~~~~~~~~~~~~~~~~~~~~~~~~~~~~~~~~~~~~\nonumber
\\=\frac{d}{dr}\left[\frac{-2r+6m~erf(\frac{r}{2\sqrt{\theta}})
-3c_1-\frac{6mr}{\sqrt{\pi \theta}}e^{-\frac{r^2}{4\theta}}}{8\pi
r^3}\right],
\end{eqnarray}

\begin{eqnarray}
F_{a}=\frac{2}{r}(p_{t}-p_{r})
~~~~~~~~~~~~~~~~~~~~~~~~~~~~~~~~~~\nonumber
\\ =\frac{2}{r}\left[\frac{-\sqrt{\pi}\theta^{2}
+6mr^{2}\sqrt{\theta}e^{-\frac{r^2}{4\theta}}}{8(r\theta)^{2}(r\theta)^{\frac{3}{2}}}
\right.  \nonumber \\
\left.-\frac{{-2r+6m~erf(\frac{r^2}{2\sqrt{\theta}})-3c_{1}-\frac{6mr}
{\sqrt{\pi \theta}}}e^{-\frac{r^2}{4\theta}}}{8\pi r^3}\right].
\end{eqnarray}

\small\begin{figure}
    \begin{center}
\includegraphics[scale=.4]{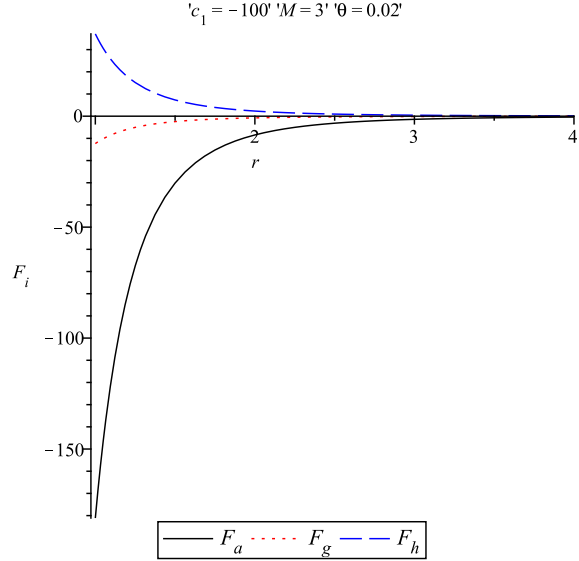}
    \end{center}
    \caption{A plot for the three different forces acting on the fluid elements in static equilibrium
against $r$. It is to be noted from the figure that due to the
joint action of $F_g$ and $F_a$ against $F_h$ the system becomes
stable at $4D$}
    \label{rf}
\end{figure}\normalsize

Even though the differential Eq. (49) for $F_h$ has not been
worked out, yet one can draw information from Eq. (47) to have a
primary conclusion. By using Maple we actually are able to plot
without writing the derivative in analytical form. Thus, as far as
the Fig. 2 is concerned, we observe that $F_a$ is the most
dominant factor whereas the least one is $F_g$. Moreover, at lower
dimension the joint action of $F_g$ and $F_a$ is much more than
$F_h$ so that the system becomes unstable. On the other hand, as
we approach towards $4D$ they balance each other and thus make a
stable configuration.

\section{Concluding Remarks}
The analysis done in the foregoing section immediately indicates
that at $4$-dimension only one can get a stable configuration for
any spherically symmetric stellar system as such higher dimension
becomes untenable as far as the stability of a system is
concerned. In a study on higher dimensional framework of
noncommutative geometry Farook et al. \cite{Rahaman2012b}, replace
pointlike structures with smeared objects and have found that
wormhole solutions exist in the usual four, as well as in five
dimensions also (only in a very restricted region), but they do
not exist in higher-dimensional spacetime.

However, it is now an open question whether the above remark is
true for a noncommutative geometry admitting conformal Killing
vectors or in any geometry this becomes feasible. In this
connection we note that Farook et al. \cite{Rahaman2009} studied a
generalized Schwarzschild spacetime with higher dimensions and
performed a survey whether higher dimensional Schwarzschild
spacetime is compatible with some of the solar system phenomena.
As a sample test they examined four well known solar system
effects, viz., (1) Perihelion shift, (2) Bending of light, (3)
Gravitational redshift, and (4) Gravitational time delay. It has
been shown by them that under a $N$-dimensional solutions of
Schwarzschild type very narrow class of metrics the results
related to all these physical phenomena are mostly incompatible
with the higher dimensional version of general relativity.

In this regard a special point is to mention that in the present
paper we deal only with spherical distributions. Therefore a
obvious question may arise: what about possible departures from
sphericity. It  is well known that multipoles play relevant roles
as can be noticed in several cases of practical interests, ranging
from tests of fundamental physics in the field of stars, planets
with crafts, to black holes and stellar systems used to probe
their
properties~\cite{Hartle1968,Laarakkers1999,Shibata2003,Iorio2006,Iorio2011a,Iorio2011b,Boshkayev2012,Renzetti2012,Iorio2013}.
So, it would be important to see what happens to such kind of
non-spherical distributions in more or less than 4-dimensions is a
pertinent issue which will be the subject of further researches in
our future projects.

Same remarks holds also for the influence of the rotation in a
Kerr/Lense-Thirring-like way: does the angular momentum of the
source influence the results, and, if so, to what extent? We note
that there are several such investigations in different
context~\cite{Barker1974,Ashby1993,Iorio2001,Iorio2012}.
Therefore, this problem may also be an issue of another future
study.

As a final comment we would like to add here that a much more deep
and comprehensive studies are required before offering a
concluding remark in connection to stability of a spherically
symmetric fluid distribution at $4D$ only.

\subsection*{Acknowledgments}
FR, AP, NA and SR are thankful to the Inter-University Centre for
Astronomy and Astrophysics (IUCAA), Pune, India for providing
research facilities. FR is also grateful to UGC, India for
financial support under its Research Award Scheme. We all express
our grateful thanks to the anonymous referee for the several
suggestions which have enabled us to improve the manuscript
substantially.

\end{document}